\documentstyle[11pt,epsfig]{article}

\begin{document}

\centerline{ Nuclear Instruments and Methods in Physics
Research A, 484 (2002) 95-101}
\bigskip
\bigskip
\bigskip

\centerline{\bf Electro-Mechanical Resonant Magnetic
Field Sensor}
\medskip
\centerline{Alexander B. Temnykh$^1$ and Richard
V. E. Lovelace$^2$}
\medskip
\centerline{1. 
Wilson Laboratory, Cornell University, Ithaca, NY 14853;
abt6@cornell.edu} 

\centerline{2. Department of Astronomy, Cornell
University, Ithaca, NY 14853;  RVL1@cornell.edu}
\bigskip

\section*{Abstract}
    We describe a new type of
 magnetic field sensor which is
termed an Electro-Mechanical Resonant Sensor
(EMRS).
    The key part  of this  sensor is a small
conductive elastic element with low damping
rate and therefore a high Q fundamental mode
of frequency $f_1$.
   An AC current is driven
through the elastic element which, in the
presence of a magnetic field,  causes an AC
force on the element.
   When the frequency of the
AC current matches the resonant frequency of
the element,  maximum vibration of the element
occurs and this can be measured precisely by
optical means.

          We have built and tested
a model sensor of this type using  for the
elastic element  a length of
copper  wire of diameter 0.030 mm  formed
into a loop shape.
  The wire motion was measured using
a light emitting diode photo-transistor assembly.
   This  sensor demonstrated a sensitivity
better than $0.001$G  for an applied magnetic
field of $ \sim 1$G
and a good selectivity for the magnetic
field direction.
   The  sensitivity can be easily
improved by  a factor of $\sim 10 - 100$ by more
sensitive measurement of the elastic element
motion and by having the element in vacuum to
reduce the drag force.

\section{Introduction}

     There are several types of magnetic field
sensors in common use.
   Included are compass
needles, Hall probes, flux-gate
magnetometers, and SQUIDs (superconductive
quantum interference devices).
   The compass needle was discovered
$\sim 2000$ years ago by the Chinese.
  Other types of sensors were
developed relatively  recently.
   The field of  application
of the different sensors clearly
depends on the required accuracy,
sensitivity, and expense.
  A compass needle is simple
and does not require
electric power or circuits but
it indicates only the field direction.
    Hall probes are more sophisticated
devices which can measure fields over
a large range of  field strengths.
   They are simple to use, but
have problems related to
baseline drift and high sensitivity to
ambient temperature
changes, and they tend to have noisy signals.
    Their typical
resolution  is
$\sim 10^{-3}$, but with
  special precautions it can reach  $5 \times
10^{-4}$.   
    Flux-gate magnetometers are
versatile and sensitive but require 
sophisticated signal
processing which tends to limit the
frequency band of their response.
    The most advanced type of magnetic
field sensors are SQUIDs
   (super-conductive quantum interference
devices).
   These probes can measure magnetic fields
with extremely high precision, but they require
liquid nitrogen or liquid helium to 
operate and sophisticated
electronics.
    That  makes this type of probe
expensive and  limits its range of
application.

In this paper we  discuss a new type of
inexpensive  magnetic field sensor which is
depicted in Figure 1.
  The sensor is a highly miniaturized
version of the vibrating wire field
measuring technique (Temnykh 1997).
   In \S 2 we briefly consider the theory
and in \S 3 we give the results of measurements
on a model EMR sensor.
    In \S 4 we give a summary.

\section{Theory}

    Here we briefly consider properties
of an elastic loop sketched
in Figure 2.
     Assume that both ends of the loop, $A$ and $B$,
are fixed.
    A torque $T$ applied to the loop will
cause it to tilt through an angle $\alpha$
around horizontal $x-$axis, where
\begin{equation}
  T  = \frac{\pi  S d^4}{16L}~ \alpha~.
  \label{Torque}
 \end{equation}
where $S$ is the shear modulus of the wire,
$d$  the wire
diameter,
$L$ is the ``arm'' length indicated in
Figure 2 (see for example Brekhovskikh
\& Goncharov 1993).
  For copper $S\approx 0.42 \times 10^{11}$Pa,
with  ${\rm Pa }\equiv {\rm N}/{\rm m}^2$.

   Consider now a DC current $I$ flowing
through the loop.
     The Lorentz force between the
current in the
loop and the vertical magnetic field $B_y$
gives a torque about the $x-$axis  approximately
  \begin{equation}
  T_I =  B_y I R^2 \left(\pi+\frac{2H}{R}-4\right)~,
  \label{Moment}
 \end{equation}
where $H$ and $R$ are defined  in Figure 2.
    In the equilibrium,
$T_I=(\pi S d^4/16L)\alpha$, so that
\begin{equation}
  \alpha  = \frac{16}{\pi}
\left(\pi+\frac{2H}{R}-4\right)
\frac{LR^2}{Sd^4}B_yI_{dc}~.
  \label{Angle}
 \end{equation}
   The displacement  of the upper point
of loop in $z-$direction is
  \begin{equation}
  \delta z = \alpha H  =
\frac{16}{\pi} \left(\pi+\frac{2H}{R}-4\right)
\frac{LHR^2}{Sd^4}B_yI_{dc} ~.
 \label{displacement}
 \end{equation}
For the loop parameters indicated
in the caption of Figure 2,
and $B_y = 110$G  (used in the tests
described below), we obtain
\begin{equation}
\frac{\delta z [{\rm mm}]}{I[A]} \approx 0.294~.
\end{equation}
  The measured dependence discussed in \S 3.3
is $\delta z [{\rm mm}]/I[{\rm A}] \approx 0.35 $.
   This is consistent with the analytic
calculation if  account is taken
of the approximate equation (2) for the torque.

  The lowest frequency of vibration $f_1$ of the
loop can be estimated as follows.
The moment of inertia of the loop about the
$x-$axes is
\begin{equation}
 I =
\frac{\pi}{4}\left[(\pi-\frac{4}{3})R^3-
2HR^2+(\pi-2)H^2R+
\frac{2}{3}H^3 \right]\rho_\ell d^2 ~.
\label{Inertia}
\end{equation}
where $\rho_\ell$ is the density of the loop.
  Thus the equation of motion for the
free vibrations of the loop is
\begin{equation}
 I ~{d^2 \alpha \over d  t^2}=
-\left({\pi S d^4 \over 16 L}\right) \alpha~.
\end{equation}
The frequency of vibration is thus
\begin{equation}
f_1=\frac{d}{4\pi} \sqrt{ { S \over \rho_\ell L}}
\left[(\pi-\frac{4}{3})R^3-2HR^2+
(\pi-2)H^2R+\frac{2}{3}H^3 \right]^{-1/2}~.
\label{Inertia}
\end{equation}
For the loop dimensions shown
in Figure 2 and for the
copper wire used
($\rho_\ell = 8.9 \times 10^3 {\rm kg/m}^3$),
equation (8) gives
$f_1 \approx280$Hz.
   This is consistent with the measured
resonance frequency discussed
in \S 3.2,   $f_1\approx 259Hz$.

   The full equation of motion for the driven
motion of the loop including the low Reynolds
number air friction is
\begin{equation}
 I ~{d^2 \alpha \over d  t^2}=
-\left({\pi S d^4 \over 16 L}\right) \alpha
- K {d \alpha \over dt}
+T_I(t)~.
\end{equation}
  Here, $T_I(t)$ is given by equation (2)
with the current $I(t)$ a function of time,
and $ K \approx (4 \pi \eta/\Lambda)
(2H^3/3+2RH^2)$, where $\eta \approx 1.8\times 10^{-5}$
kg/(m s) the dynamic
viscosity of air.   Also, for a long cylinder,
$\Lambda =1/2 -\gamma -\ln(v_{p} d/8\nu)$
(Landau \& Lifshitz 1959), with
$\gamma \approx 0.577 $ Euler's constant,
with $v_{p}$ the peak velocity of the top of
the loop, and with
$\nu = \eta/\rho_{air}\approx 1.5\times 10^{-5}
{\rm m^2/s}$ the kinematic viscosity of air.
    Equation (9) implies that the quality
factor for the vibrations is
$Q = 2\pi f_1 I/ K$.
   For the displacement amplitudes
discussed in \S 3.2, the Reynolds numbers
$Re = v_p d/\nu$ are indeed less than
unity, and this formula gives
the prediction  $Q \approx 297$ which is larger
than the measured value of $198$ (\S 3.2).
   We believe that this difference is
due to the  approximations in the theory
of the drag coefficient $ K$.

 \section { EMR Sensor Tests}

   We built a number of models of  EMR
sensors.
   One of these are shown schematically
in Figures 1 and 2.
    The elastic element
was fabricated from a $0.030$mm diameter copper
wire formed into a loop shape diagrammed in
Figure 2.
   The fundamental mode
of vibration corresponds to the
top of the loop
moving in the $z-$direction in
Figure 2.
     To measure position of the loop
 we used a ``$\Pi$'' shaped opto-electronic
assembly H21A1 (Newark Electronics)
consisting of a light-emitting-diode
(LED) on one leg
of the assembly and a
 photo-transistor  on other.
   The
light flux detected by
photo-transistor is very sensitive to loop
position.

In the following we
give the characteristics of the EMR
sensor components and   results of the sensor
tests.

\subsection{Calibration of  Loop Position
Sensor}

   The opto-electronic detector  was
calibrated by the moving the entire  loop
using a precise micro-screw
and measuring the signal
from the photo-transistor.
   The measured
dependence is shown in
Figure~\ref{DetectorCalib}.

One can see that in the range
$0.7 -1.1 $mm the  signal from
photo-transistor is proportional to the loop
displacement.
\begin{equation}
  \frac{\delta U[{\rm mV}]}
{\delta  z[{\rm mm}] } = 421
  \label{Calib}
 \end{equation}
Here, $\delta z$ is the change
of the loop position, $\delta U$ is
the change in the detector
signal. $\delta U = 1mV$ corresponds to a current $\delta I = \delta U /R =
0.16 \mu A$ through
the external circuit resistor $R=6.35 k \Omega$.
    In subsequent measurements, the
loop position was adjusted to be in the middle
of the range of linear dependence.

\subsection{Elastic Element Resonance Response}

    Important
characteristics of EMR probe are the
fundamental resonance frequency $f_1$ and
the quality factor $Q$.
    We measured these parameters
by driving an AC current with
various frequencies through the element and
measuring amplitude of the AC signal from
photo-transistor.
    In Figure~\ref{Resonance}
the measured amplitude ( RMS of AC voltage) is plotted as a function
of frequency $f$ of the AC current.

The data  was fitted to the
resonance formula
\begin{equation} A  = \frac{A_0}{ \sqrt
{(f^2-f_1^2)^2+f^2 f_1^2 /Q^2}},
\label{ResFormula}
 \end{equation}
where, $f_1$ is the
resonance frequency, $Q$ is
the quality factor, and $A_0$
is a constant.  This expression follows
from equation (9).
   The fit gave:
$f_1 = 259$Hz and   $Q = 198$.
Note that according to
calibration~\ref{Calib} the maximum of $16$mV
RMS of AC signal seen at resonance  on
Figure~\ref{Resonance} corresponds to $\pm
0.053$mm of amplitude of vibration of
the top of the loop.
  This small amplitude indicates that the
optical detector was  operating in the
linear region of Figure 3.
   In this test the
AC current amplitude trough the element was
$170$mA and the magnetic field was
$\sim 0.5$G.

\subsection{Elastic Element Static Test}

  This
test was done to measure the static properties
of the elastic element.
   A triangular  AC
current of  low frequency of $1$Hz  was
driven through the element.
   A $110$G
vertical magnetic field  was
imposed at the EMR probe location  by a
permanent magnet.
   The Lorentz force
between the magnetic field and current flowing
through the probe  caused the loop displacement.

Figure~\ref{StatDefla} shows the current through
the element, $I_{el}$, and photo-transistor
signal as a function of time. 
   One can see the
triangular wave current with $1sec$ period and
photo-transistor signal with similar form.
   In Figure ~\ref{StatDeflb} the signal is
plotted as function of current. 
   The right-hand  vertical
scale  shows the element
displacement, $z$,  calculated from the signal
by using calibration~\ref{Calib}. The data
indicates a linear dependence of element
displacement on current,
 \begin{equation}
\frac{\delta z [{\rm mm}]}
{\delta I[{\rm A}] }\approx  0.365~,
 \end{equation}
for the magnetic
field of $110$G.

\subsection{EMR Sensor Calibration and Comparison
with a Hall Probe}

    In this test a Hall probe was placed very
close to the EMR probe.
    The Hall probe orientation
was accurately adjusted so  that it sensed only
the vertical component of the magnetic field, 
which is the  component which causes
the vibration of elastic element in EMR probe.
    A sinusoidal  AC
current of peak amplitude $85$mA  at the
resonance frequency $259$Hz was driven
through the element.
   The imposed magnetic field was
created by a small permanent magnet.
    The field
strength was varied by accurately moving the
permanent magnet relative to the fixed probes.
    Figure \ref{EMRandHallReading} shows
the field measured by the Hall probe and the
signal from the EMR sensor
  as a function of the test magnet position.

The data shown in
Figure~\ref{EMRandHallReading} was used to
calibrate the EMR probe.
   Figure~\ref{EMRvsHall} shows a plot
of the  EMR probe signal (RMS of AC voltage) 
versus  magnetic
field strength measured with Hall probe and
fitted it with linear dependence.
   This fit gives
\begin{equation}
\frac{\delta B[{\rm G}]}{\delta U_{AC}[{\rm mV}]} \approx
0.0518~.
\label{EMR_calibration}
\end{equation}
That is, a $1$mV change of the EMR probe
signal  indicates a $52$mG change of the magnetic
field strength.

Note that because the size of the test
magnet   was  much smaller than the
distance  between probes and the magnet, the
dependence of the magnetic field strength
on the position $p$
can be approximated by
 \begin{equation}
 B(p) = C_1 + \frac
{C_2}{(p-C_3)^3 }~.
\label{fit3}
 \end{equation}
Here, $C_1$ is a parameter which  represents
either the background field
or the $zero$ drift for the
Hall probe.
  The parameter $C_2$ is proportional to
the testing magnet magnetic moment,
  The parameter $C_3$  is set by the
location of the probe in coordinate system used
to define  testing magnet position.
    The
measurements with the Hall and EMR probes  were
fitted with the theoretical
dependence~\ref{fit3} using $C_1$, $C_2$ and
$C_3$ as a free parameters.
   The residual
between measured data and theoretical fit for
both probes is shown in
Figure~\ref{EMR_Hall_Comparison}.

 For the EMR probe the difference
(measurement - fit) was converted into magnetic
field strength using equation (\ref{EMR_calibration}).
  Note that at each point, the
signal from EMR probe was measured several times.
Bars shown for the EMR sensor data represent
$1\sigma$  errors found from a statistical
analysis of the measurements.

   We can now compare the difference between
measurement and theoretical fit for
the EMR sensor and that for the
Hall probe.
   For Hall probe RMS the residual
between measured data and theoretical fit  is
$2.9 \times 10^{-3}$G, which is
consists with  the  probe specifications.
  For our EMR sensor  the
residual is $0.45 \times 10^{-3}$G.
That is, the EMR sensor is 6 times better!

\section{Discussion}

   In future refinements,
an EMR sensor  element optimized for sensitivity
can be developed.
    The geometry
may be similar to that discussed above but
the wire can be of diameter
$d=0.010$mm 
and consist of $10$ turns around the loop
shown in Figures 1 and 2.
   Scaling of our test results using equations (~\ref{displacement})
and  (\ref{Inertia})  gives a sensitivity $5\times10^{-6}$G,
resonance  frequency  $\sim 25$Hz,  and  a peak  current amplitude
$9.4$mA where the current was scaled $ \propto d^2$.   The elastic
element can be vacuum encapsulated using nano-fabrication techniques.
  For vacuum conditions where
the mean-free-path is longer
than the dimensions of the loop, the
friction coefficient $ K$ in
equation (9) is proportional
to the gas density or pressure.
    We estimate that the
quality factor of the resonant element
can be increased to $Q \sim 10^3 - 10^4$.
    This increase in $Q$ will  increase
the sensitivity by a
factor $\sim 5-50$.
   As a result the EMR sensor can have a sensitivity of
$1 \times 10^{-6}$ to $10^{-7}$G.
     This is a few orders of magnitude more sensitivity than
a Hall probe, but of course it is not as sensitive as SQUIDS.
    Note that the measurement times at the lowest field levels need to be
$\sim 1-10$min.

\thanks{We thank Drs. M.M. Romanova
and S. Temnykh for helpful discussions.}

\section*{References}

\noindent Brekhovskikh, L. M., \& Goncharov, V. 1993, {\it Mechanics of Continua and 
Wave Dynamics} (Berlin: Springer)

\medskip

\noindent Landau, L.D., \& Lifshitz, E.M. 1959,
{\it Fluid Mechanics} (Pergamon Press: London), p. 68

\medskip

\noindent Newark Electronics 2000, Catalog 118, p. 586.

\medskip

\noindent Temnykh, A. 1997, Nuclear Instruments \&
Methods in Physics Research, 399, 185

\begin{figure}[p]

{\Large\bf Figure Captions}

\caption{\label{EMR_Imagel}
Three dimensional view of
sample EMR sensor.
   The wires marked
(a) and (b) connect to the conducting
loop in the center of the figure. The ``$\Pi$''
shaped assembly with light-emitting-diode (LED) on one leg and
a photo-transistor on other  is used as the loop position
detector. The straight white segment above the
loop represents the light beam of LED.
  The photo-transistor current  indicates the position
of the top of the loop.}

\caption{\label{Loop1}
  Schematic view of elastic
element used in tests.
  The element was  made of
copper wire of diameter $d=  0.030$mm.
  The other dimensions are $H =4$mm,
$R= 1$mm, and $L= 6.25$mm.}

\caption{\label{DetectorCalib}
    Signal from
photo-transistor (U) as a function of the loop
position.
  A least-squares
fit of a straight line through the solid
circles gives
$U[{\rm mV}]= m_0+m_1z[{\rm mm}]$ with
$m_0=-110$ and $m_1=421$. }

\caption{\label{Resonance}
   EMR sensor
resonance characteristic. The curve represent a
least-squares fit of equation (9) to
the measured points. This fit gives
 $f_1=259$Hz and $Q=198$.}
 
\caption{\label{StatDefla}
  EMR sensor static
test for an imposed magnetic field $B=110$G.
Current through the elastic element (dashed line) and
signal from photo-transistor (solid line)
as a function of time.}

\caption{\label{StatDeflb} EMR sensor static
test for an imposed magnetic field of $B=110$G.
Signal from photo-transistor (left-hand
scale)  as a function of current through the
element. Right-hand scale is the loop
 deflection $z$ in mm.  The linear fit of the
data gives: $z[{\rm mm}]= -3.96 \times 10^{-2} + 0.365
 I[{\rm A}]$. }
 
\caption{\label{EMRandHallReading}
   Signal
from EMR sensor (open circles)  and magnetic
field measured by the Hall probe (solid triangles)
as function of  test magnet position. }

\caption{\label{EMRvsHall}
  RMS of AC signal from EMR sensor
($U_{AC}$) as function of the magnetic field measured
with Hall probe ($B$) at $170$mA of peak-to-peak
current driven through the loop.
A least-squares fit gives $ U_{AC}[{\rm
mV}]=1.95 +
19.31  B[{\rm G}] $  or $ \delta B[{\rm G}]
\approx  0.0518  \delta U_{AC}[{\rm mV}]$.
}

\caption{\label{EMR_Hall_Comparison}
Residual of measured  fit for EMR
sensor (circles) and the Hall
probe (diamonds).
The Hall probe RMS = $2.9 \times
10^{-3}$G, whereas the EMRS RMS = $0.45 \times
10^{-3}$G.  }
  
\end{figure}

\end{document}